\documentclass[traditabstract]{aa}
\usepackage{lscape}
\usepackage{longtable}
\usepackage{graphicx}
\usepackage{txfonts}
\begin{document}
   \title{\emph{Swift}/XRT counterparts to unassociated \emph{Fermi} high-energy LAT sources}


    \titlerunning{\emph{Swift}/XRT counterparts to unassociated \emph{Fermi} high-energy LAT sources}
\authorrunning{R.~ Landi et al.}

   \author{R.~ Landi \inst{1} \and L.~Bassani \inst{1} \and J. B. Stephen \inst{1}
\and N.~Masetti \inst{1} \and A.~Malizia \inst{1} \and P. Ubertini \inst{2}}
   \offprints{landi@iasfbo.inaf.it}
\institute{INAF/IASF Bologna, via Piero Gobetti 101, I--40129 Bologna, Italy
\and
INAF/IAPS Rome, Via Fosso del Cavaliere 100, I--00133 Rome, Italy}

   \date{Received  / accepted}

\abstract{
We report the results from our analysis of a large set of archival data acquired with the X-ray 
telescope (XRT) onboard \emph{Swift}, covering the sky 
region surrounding objects from the first \emph{Fermi} Large Area Telescope (LAT)
catalogue of high-energy sources (1FHL), which still lack an 
association. Of the 23 regions analysed, ten did not show any evidence of X-ray emission, but 13 were 
characterised by the presence of one or more objects emitting in the 0.3--10 keV band. Only in a couple of 
cases is the X-ray counterpart located outside the \emph{Fermi} positional uncertainty, while in all other 
cases the associations found are compatible with the high-energy error ellipses. All counterparts we found 
have been studied in detail by means of a multi-waveband approach to evaluate their nature or class;
in most cases, we have been able to propose a likely or possible association except for one \emph{Fermi} 
source whose nature remains doubtful at the moment. The majority of the likely associations are 
extragalactic in nature, most probably blazars of the BL Lac type.
}


\maketitle

\section{Introduction}

A key strategic objective of the \emph{Fermi} mission is to perform a survey of the sky at various 
gamma-ray energies. Recently, the first \emph{Fermi}/LAT
catalogue of high-energy sources (1FHL) has been published 
(Ackermann et al. 2013), listing 514 objects detected above 10 GeV. The main motivation behind this 
catalogue was to find the hardest gamma-ray sources in the sky and to get a sample of objects that are 
good candidates for detection at TeV energies. The vast majority of the sources reported therein could be 
immediately associated with known objects (449 or 87\% of the sample): approximately 75\% with 
active galactic nuclei (AGNs), mostly 
blazars, while Galactic sources, like pulsars, pulsar wind nebulae (PWNs), supernova remnants 
(SNRs), high-mass binaries, and star-forming regions,
collectively represent 10\% of the sample. The percentage of unassociated sources is less than 14\% 
corresponding to 71 objects. The third \emph{Fermi}/LAT catalogue (3FGL, Acero et al. 2015), 
which was subsequently 
published, contains most of these unassociated 1FHL sources except for 13 objects that are missing. 
Considering the associations discussed in the 3FGL catalogue (six objects) and our follow-up work 
(26 objects analysed by Landi et al. (2015a);(2015b)), the set of sources for which no counterpart
has been found found contains at this stage 39 entries.

Identification of these objects is a primary objective of the mission,
but it has been made difficult by the relatively large positional uncertainty (usually around a few 
arcminutes) of the \emph{Fermi}/LAT instrument. 
Analysis of X-ray data can 
be a useful tool for restricting the positional uncertainty of the \emph{Fermi} objects and 
facilitating the identification process. The list of 
likely counterparts that are found can then be studied in detail (for example using multi-waveband data) 
to identify those highly nusual objects with the parameters that might be expected to produce gamma-rays.
Optical spectroscopy can finally be performed to confirm the association and classify the gamma-ray 
source.

Here we use data collected with the XRT instrument onboard the satellite \emph{Swift} (Gehrels et al. 
2004) to study a number of unassociated \emph{Fermi} 1FHL sources. We do this by cross-correlating the list 
of 39 1FHL objects that lack a counterpart, with all the XRT pointings covering the 
\emph{Fermi}/LAT 
positional uncertainty. Analysis of these data allows us to propose the likely or possible  
counterparts to another 12 \emph{Fermi} high-energy objects; the nature of each likely or possible 
counterpart is studied by means of a multi-waveband approach and is generally found to conform to 
the nature of blazars of the BL Lac type.

\section{\emph{Swift}/XRT follow-up observations}

We took all 39 sources in the 1FHL catalogue that still lack an association and class. The source 
positions were 
then cross-correlated with the \emph{Swift}/XRT archival data, and we only considered distances between the 
\emph{Fermi} source and XRT pointing positions below 9 arcmin in order to optimise coverage of the 
high-energy error ellipses. This cross-correlation method led to selecting a sample of 23 
1FHL sources for which XRT data are available. The log of all X-ray observations analysed in this 
work is given in Table~\ref{tab1}, where we report for each individual \emph{Fermi} source, the XRT 
observation ID, the date and the exposure of the XRT pointings available up to March 15, 2015.

XRT data were reduced using the XRTDAS standard data pipeline package (\textsc{xrtpipeline} v. 
0.12.9) to produce PC-mode screened event files. For each \emph{Fermi} source, we summed together all the 
available XRT pointings using \textsc{XSELECT v. 2.4c} to enhance the signal-to-noise ratio and thus allow 
the detection of possible counterparts. The XRT images in the 0.3--10 keV energy band were obtained and 
analysed by means of the software package \textsc{XIMAGE} v. 4.5.1; some of the XRT images are shown in 
Figures from 1 to 4 and discussed in dedicated sections.

From the XRT data we did not find evidence of soft X-rays for ten 
sources in our sample and these are listed in Table~\ref{tab2}, where we also quote an upper limit on the 
count rate in the 0.3--10 keV energy band and the Galactic neutral hydrogen column density in the 
source direction.
  
For each of the remaining objects, XRT 
detected one or more sources\footnote{We have taken into account all XRT detections above 
2.5$\sigma$ confidence level.} in the region containing or surrounding the \emph{Fermi} positional 
uncertainty. These cases are listed in Table~\ref{tab3} and discussed in the following sections.
We also provide further X-ray information in Table~\ref{tab4}: spectral parameters 
only for those sources for 
which the quality of the XRT data is good enough to perform a basic spectral analysis, i.e. a simple 
power law passing through Galactic absorption. In the remaining 
cases, we only estimate the 0.3--10 keV count rate and the Galactic neutral hydrogen column 
density in the source direction.

For the spectral analysis, source events were extracted within a circular region with a radius of 20 pixels
(1 pixel $\sim$ 2.36 arcsec) centred on the source position, while
background events were extracted from a source-free region close to the X-ray source of interest.
The spectra were obtained from the corresponding event files using the \textsc{XSELECT v. 2.4c}
software and binned using \textsc{grppha} in an appropriate way, so that the $\chi^{2}$ statistic could
be applied. We used version v.014 of the response matrices and created individual ancillary
response files \emph{arf} using \textsc{xrtmkarf v. 0.6.0}.

Finally, we point out that only four objects from Table~\ref{tab3} (1FHL J0639.6--1244, 1FHL J1240.4--7150, 
1FHL J1507.0--6223, and 1FHL J1856.9+0252) are located close to the Galactic plane (i.e. below 10 degrees 
in Galactic latitude), suggesting that the remaining nine sources are most likely extragalactic.

\section{\textbf{Individual source studies}}

In the following, we provide detailed information available in the literature and in various archives 
on each individual association with the \emph{Fermi} high-energy sources found in this work. Besides the 
X-ray information, 
we also checked the radio, infrared (by exploiting the Wide-field Infrared Survey Explorer 
(WISE, Wright et al. (2010)), and optical characteristics of each possible counterpart in order to 
understand its nature in more detail and to assess the likelihood of its association with the \emph{Fermi} 
object. More specifically, in Table~\ref{tab3} we report, for each \emph{Fermi} high-energy source for 
which we find one or more X-ray counterparts, the name, the number of X-ray sources detected with XRT in 
the corresponding fields (and next to it a 
reference number used in the WISE colour-colour plot (see Figure 5), their coordinates and error radius, 
the name of the WISE counterparts, its magnitudes ($W1=3.5$, $W2=4.6$, $W3=12$, $W4=22$ 
micron)\footnote{available at:\\ http://vizier.u-strasbg.fr/viz-bin/VizieR?-source=II\%2F311.}, and the 
likelihood of association with the corresponding gamma-ray source. It is evident from Table~\ref{tab3} 
that the XRT 
location accuracy allows a significant reduction in the \emph{Fermi} positional uncertainty, thus making 
the search for possible optical counterparts much easier. 

In most cases, the soft X-ray error radius is 
lower than or equal to six arcseconds, i.e. sufficiently small to highlight only one optical counterpart. 
Also X-ray information as reported in Table~\ref{tab4} can be used to characterise each source, 
in particular the count rate, and the 2--10 keV flux can help to distinguish multiple 
counterparts.
We used the WISE colours as discussed by Massaro et al. (2013a) to test the possible blazar nature of each 
source. These authors found that in the $W2-W3$ versus $W1-W2$ colour-colour plot the positions of 
gamma-ray emitting blazars are all within a well-defined region known as the ``Blazar Strip''. Therefore, 
objects with IR colours compatible with this strip could be AGN of the blazar type, hence be an even more 
likely counterpart to the \emph{Fermi} high-energy sources. For the radio information, we either use data 
collected at 20 cm by the NRAO VLA Sky Survey (or NVSS, Condon et al. 1998), at 36 cm by the Sydney 
University Molonglo Sky Survey (or SUMSS, Mauch et al. 2003) and at 92 cm by the Westerbork Northern Sky 
Survey (or WENSS, Rengelink et al. 1997) or Westerbork In the Southern Hemishpere (or WISH, De Breuck et 
al. 2002).

We also consulted the results of Schinzel et al. (2015), who performed an All Sky survey 
between 5 (6) and 9 (3.3) GHz (cm) of sky areas surrounding unidentified objects listed in the second
\emph{Fermi}/LAT catalogue (2FGL, Nolan et al. 2012) in the search for new AGN associations with compact 
radio sources. Nine of our regions were covered by this survey, but only five were found to have a radio 
detection compatible with our proposed X-ray association. They are discussed in each individual 
section.

In the following, sources will be divided into three groups: those with a single counterpart, 
those with multiple associations within or at the border of the \emph{Fermi} positional uncertainty, 
and those with a counterpart outside the \emph{Fermi} error ellipse.

\subsection{\textbf{Single counterpart sources}}

In this section, we discuss those cases in which we find only one X-ray source inside the \emph{Fermi} 
positional uncertainty. This information is used with the overall properties of the X-ray detection 
found to evaluate the likelihood of the association with the \emph{Fermi} high-energy object.

\begin{figure}[h]
\centering
\includegraphics[width=1.\linewidth,angle=0]{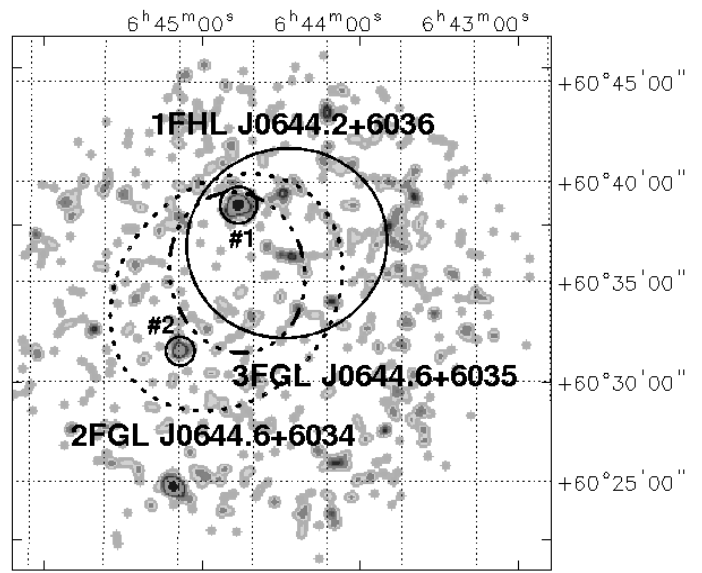}
\caption{XRT 0.3--10 keV image of the region surrounding 1FHL J0644.2+6036 
(black ellipse)/2FGL J0644.6+6034 (black-dashed ellipse)/3FGL J0644.6+6035 (black-dashed-dotted ellipse).
Source \#1 is the only XRT detection compatible with the 1FHL J0644.2+6036 positional uncertainty.
Source \#2 represents the counterpart
to 2FGL J0644.6+6034 discussed by Massaro et al. (2013b) and Paggi et al. (2014). A third source 
located outside both \emph{Fermi} error ellipses is neither numbered nor discussed in the text.} 
\label{f1}
\end{figure}

\subsubsection{1FHL J0644.2+6036}

The XRT counterpart found in this work (source \#1 in Figure~\ref{f1}) is listed both in the NVSS (NVSS 
J064435+603849) and in the WENSS (B0640.0+6041) catalogues with 20 and 92 cm flux densities of 
$\sim$33.5 and 
$\sim$83 mJy, respectively (spectral index $\alpha= 0.6$)\footnote{Here and in the 
following, we use 
$S\propto\nu^{\alpha}$, where $S$ is the flux density, $\nu$ is the frequency, and $\alpha$ the spectral 
index.}. 
The WISE colours are $W2-W3=1.97$ and $W1-W2=0.64$, making this object compatible with 
the blazar strip. The high-energy \emph{Fermi} source is also reported in the second /third\emph{Fermi}/LAT
catalogues as 2FGL J0644.6+6034/3FGL J0644.6+6035, an unidentified source studied by various authors. 
Massaro et al. (2013b) propose WISE J064459.38+603131.7 as the possible IR counterpart to 2FGL 
J0644.6+6034, while Paggi et al. (2013) find that this WISE source is also detected by \emph{Swift}/XRT. 
Recently, Paggi et al. (2014) have provided the optical spectrum of this possible counterpart, finding 
that it 
could be a weak emission line quasar at $z=0.3582$. As is evident in Figure~\ref{f1}, the positional 
uncertainties of 2FGL J0644.6+6034, 3FGL J0644.6+6035 
and 1FHL J0644.2+6036 are partly overlapping. The error ellipse of 2FGL J0644.6+6034, 
apart from the source discussed by Massaro et al. (2013b) and Paggi et al. (2014) (source \#2), contains 
also source \#1 found in this work, while the error ellipse of 3FGL J0644.6+6035 contains only this XRT 
detection, which incidentally is also the brightest of the two X-ray emitters. The lack of any radio 
detection so far for source \#2, combined with the above-mentioned characteristics of source \#1, raise 
some questions about the reality of the association between WISE J064459.38+603131.7 and 2FGL J0644.6+6034,
and suggest instead that source \#1 is the only putative counterpart to the 2FGL, 3FGL, and 1FHL 
\emph{Fermi} object.
As a confirmation of our findings, we note that source \#1 is also reported in the survey work 
by Schinzel et al. (2015) as a flat-spectrum compact radio source detected with a 6 cm flux 
density of $\sim$8.2 mJy.

\subsubsection{1FHL J1115.0--0701}

This high-energy \emph{Fermi} source has a counterpart in the still unclassified object 3FGL J1115.0--0701. 
The only XRT detection found is inside and at the border of the 1FHL and 3FGL error ellipse, respectively. 
The source has no radio counterpart or WISE colours ($W2-W3=1.825$ and $W1-W2=0.069$) that locate it 
outside the blazar strip. Both indications cast doubts on the association between this XRT detection and 
the \emph{Fermi} high-energy source, which therefore we consider at this stage uncertain until optical 
follow-up observations provide a reliable classification of this X-ray source.

\subsubsection{1FHL J1129.2-7759}

This XRT source has a counterpart in the \emph{ROSAT} Bright 
(1RXS J113029.0--780108, 7 arcsec error radius) source 
catalogue (Voges et al. 1999), as well as in a \emph{XMM-Newton} Slew survey catalogue (Saxton et 
al. 2008) (XMMSL1 J113031.6--780112, 3 arcsec error radius); indeed, its X-ray flux, listed in 
Table~\ref{tab4}, is relatively high.

This X-ray source is most likely associated with the radio source PMN J1130--7801. 
The PMN object has a 36 cm 
flux density of $\sim$136 mJy and a flat ($\alpha=-0.4$) radio spectrum from 3.5 to 6 cm (McConnell et 
al. 2012). The 
source is also reported as a WISE object with the following IR colours: $W2-W3=1.93$ and $W1-W2=0.6$, 
which are fully compatible with the blazar strip. The source is discussed in a study of the Chamaeleon 
star-forming region (Alcala et al. 1995): it has been spectroscopically investigated, but unfortunately 
with no success. We also note that this X-ray/radio source falls within the positional uncertainty of the 
3FGL counterpart (3FGL J1130.7--7800) of this 1FHL source.

\subsubsection{1FHL J1223.3+7953}

This X-ray source is listed in the Two Micron All Sky Survey Extended sources catalogue (Skrutskie et al. 
2006) as 2MASX J12235831+7953279 and was classified as a galaxy in NED. The source is also found in both 
the NVSS 
(NVSS J122358+795329) and WENSS (WN 1222.0+8010) catalogues with 20 and 92 cm flux densities of 
$\sim$31.5 and $\sim$43 mJy, respectively.
A power-law fit between these two frequencies indicates a rather flat ($\alpha 
\sim 0.2$) radio spectrum. 
This is confirmed by the findings of Schinzel et al. (2015), who report 
parsec scale radio emission with a flat spectrum for this X-ray source.
The WISE colours ($W2-W3=1.92$ and $W1-W2=0.48$) place the source well inside 
the blazar strip. This object is also listed in the Galaxy Evolution Explorer All-Sky Survey (GALEX, 
Bianchi et al. (2011)) as GALEXASC J122357.56+795328.7 with magnitudes of 24.6 and 22.3 in the far- and 
near-UV. Massaro et al. (2015) have recently identified it as a BL Lac of unknown redshift. We 
note that this 
1FHL source is listed as 3FGL J1222.7+7952 in the third \emph{Fermi} catalogue, and the XRT source lies at 
the border of its positional uncertainty confirming 2MASX J12235831+7953279 as the counterpart to both 
gamma-ray detections.

\subsubsection{1FHL J1240.4--6150}

The 3FGL association for this source is 3FGL J1240.3--7149. The only XRT source found is inside the error 
ellipses of both 1FGH and 3FGL objects. It is associated with the \emph{XMM-Newton} object XMMSL1 
J124021.7--714854 (2.0 arcsec error radius) and displays a relatively high X-ray flux (see 
Table~\ref{tab4}). It has a radio counterpart (MGPS J124021--714901) that shows a 
$\sim$15.5 mJy flux density at 36 cm (Murphy et al. 2007) and a flat radio spectrum between 5 and 
9 GHz ($\alpha=-0.45$, Petrov et al. 2013). 
In this case Schinzel et al. (2015) report parsec scale radio emission 
emission with 6 and 3.3 cm flux densities of $\sim$14 and $\sim$11.2 mJy, respectively ($\alpha=0.3$) 
for this X-ray source.
The source is listed in the WISE catalogue with colours $W2-W3=1.34$ and 
$W1-W2=0.52$, which places it outside the blazar strip.

\subsubsection{1FHL J1315.7--0730}

In this case, the XRT source is associated with a relatively weak radio source (NVSS J131552--073301), 
displaying a 20 cm flux density of $\sim$23.8 mJy and a flat ($\alpha \sim -0.1$) radio spectrum 
(Petrov et al. 2013). 
Radio emission is also reported by Schinzel et al. (2015) with 6 and 3.3 cm flux densities of 
$\sim$40.2 and $\sim$38.4 mJy, respectively, confirming the flat radio spectrum. This source is 
WISE-detected
with colours $W2-W3=2.27$ and $W1-W2=0.88$, which are again compatible with 
those of gamma-ray emitting blazars. In UV the source can be associated with a GALEX object (GALEXASC 
J131553.03--073302.4), which displays variability and high polarization (Bauer et al. 2009; 
Fujiwara et al. 2013). On the basis of these peculiarities, Massaro et al. (2013b) suggest that it
could be a new gamma-ray blazar, while Hassan et al. (2013) conclude that it
might be a BL Lac object. This 1FHL source is reported as 3FGL J1315.7--0732
in the third \emph{Fermi} catalogue.
Our XRT detection lies well within the 3FGL positional uncertainty, 
thus also confirming our proposed association for the lower energy gamma-ray emitter.

\subsubsection{1FHL J1507.0--6223}

The only X-ray source detected by XRT lies on the border of the \emph{Fermi} error ellipse and has no radio 
counterpart. Furthermore, the WISE colours $W2-W3=1.17$ and $W1-W2=-0.011$ place this source well outside 
the blazar strip. In the third \emph{Fermi} catalogue, this source is reported as an unclassified source 
that is nevertheless associated with HESS J1507--622. The three error ellipses (1FHL, 3FGL, and HESS) 
overlap considerably, but the smaller positional uncertainty of the third \emph{Fermi} catalogue excludes the 
XRT object as a possible counterpart. This sky region has been studied in depth at X-rays energies using 
\emph{Chandra}, 
\emph{XMM-Newton}, and \emph{Suzaku} (Tibolla et al. 2014; Eger et al. 2015) data. These observations, which 
are by far more sensitive than ours, revealed several X-ray sources, but not the XRT one discussed here.
This lack of detection may be due either to source variability or to different exposure times. 
Of the two sources that are compatible with the restricted 3FGL error ellipse, one (source X3, Figure 1 
in Tibolla et al. 2014) is a star and the other (source S6, Figure 1 in Eger et al. 2015) is too faint 
for an in-depth analysis. 

At the present stage the most accredited association with HESS 1507--622 is a faint 
extended X-ray source thought to be a PWN; although this source is at the border of the HESS and 1FHL 
positional uncertainties, it is well outside the 3FGL ellipse, which casts some doubt on its true 
association with the MeV/TeV object. This source clearly deserves further study, but on the basis of the 
above information, we consider the association between the XRT and the \emph{Fermi} high-energy source 
unlikely.

\begin{figure*}[t!]
\centering
\includegraphics[width=0.45\linewidth,angle=0]{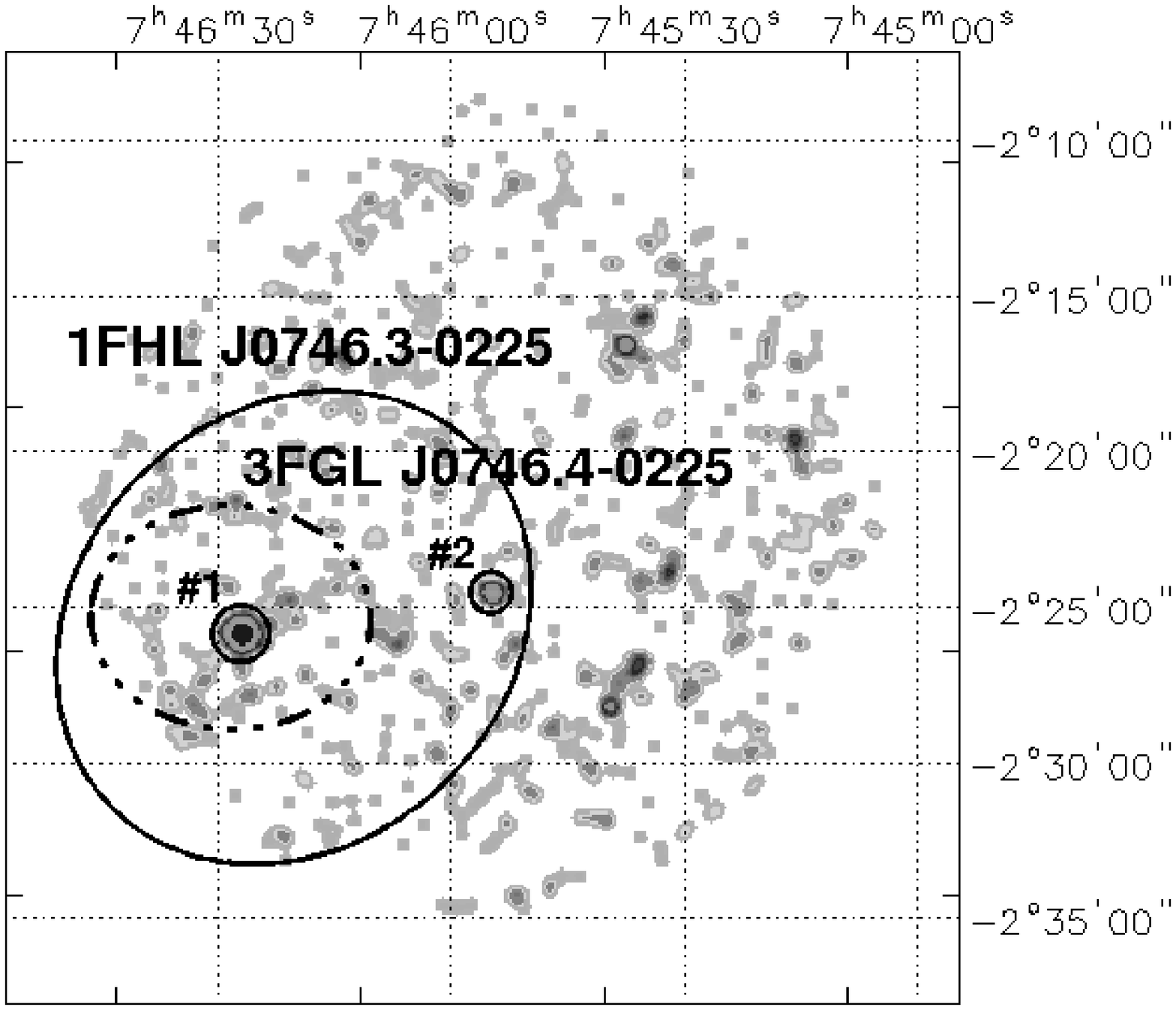}
\includegraphics[width=0.45\linewidth,angle=0]{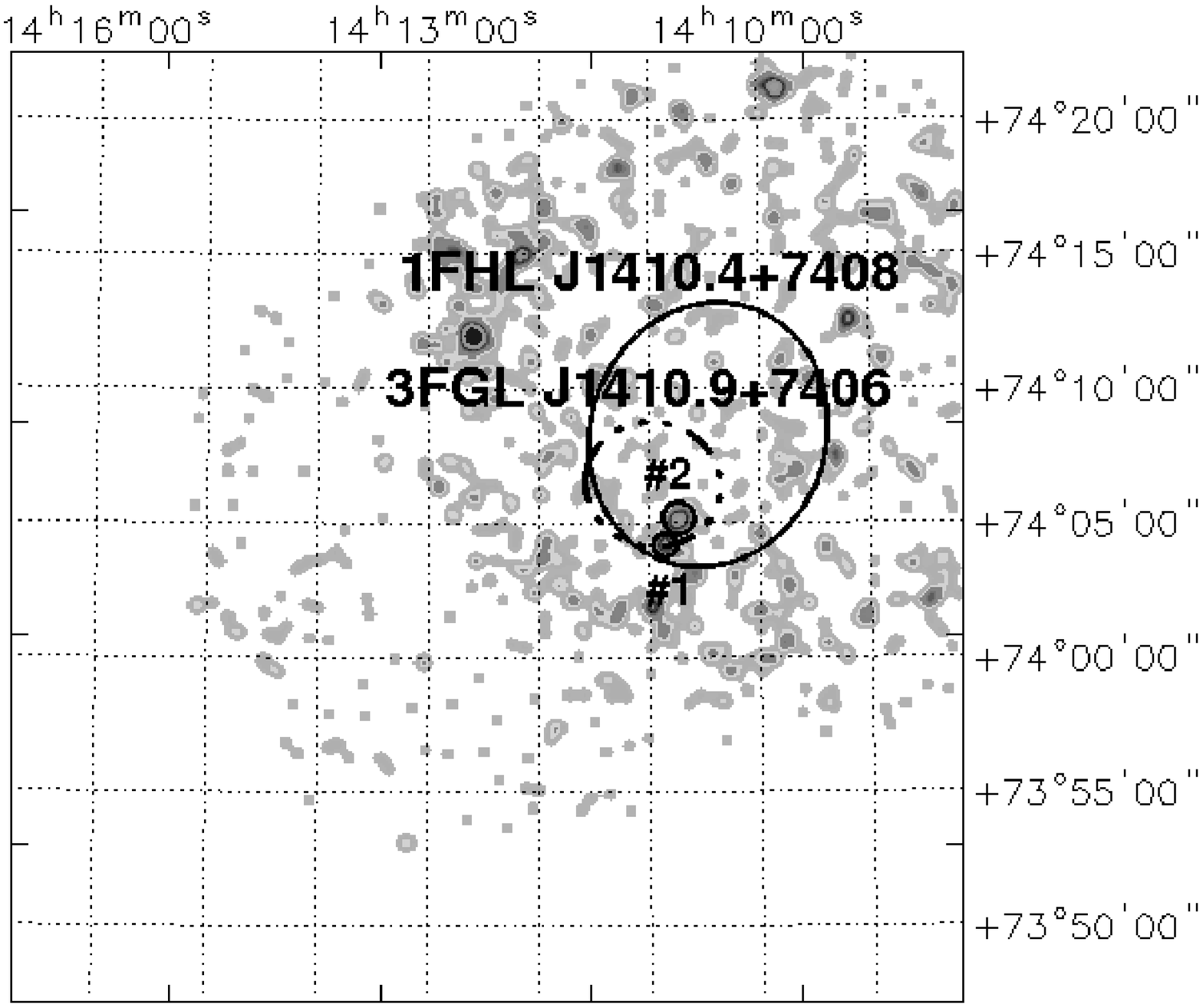}
\includegraphics[width=0.45\linewidth,angle=0]{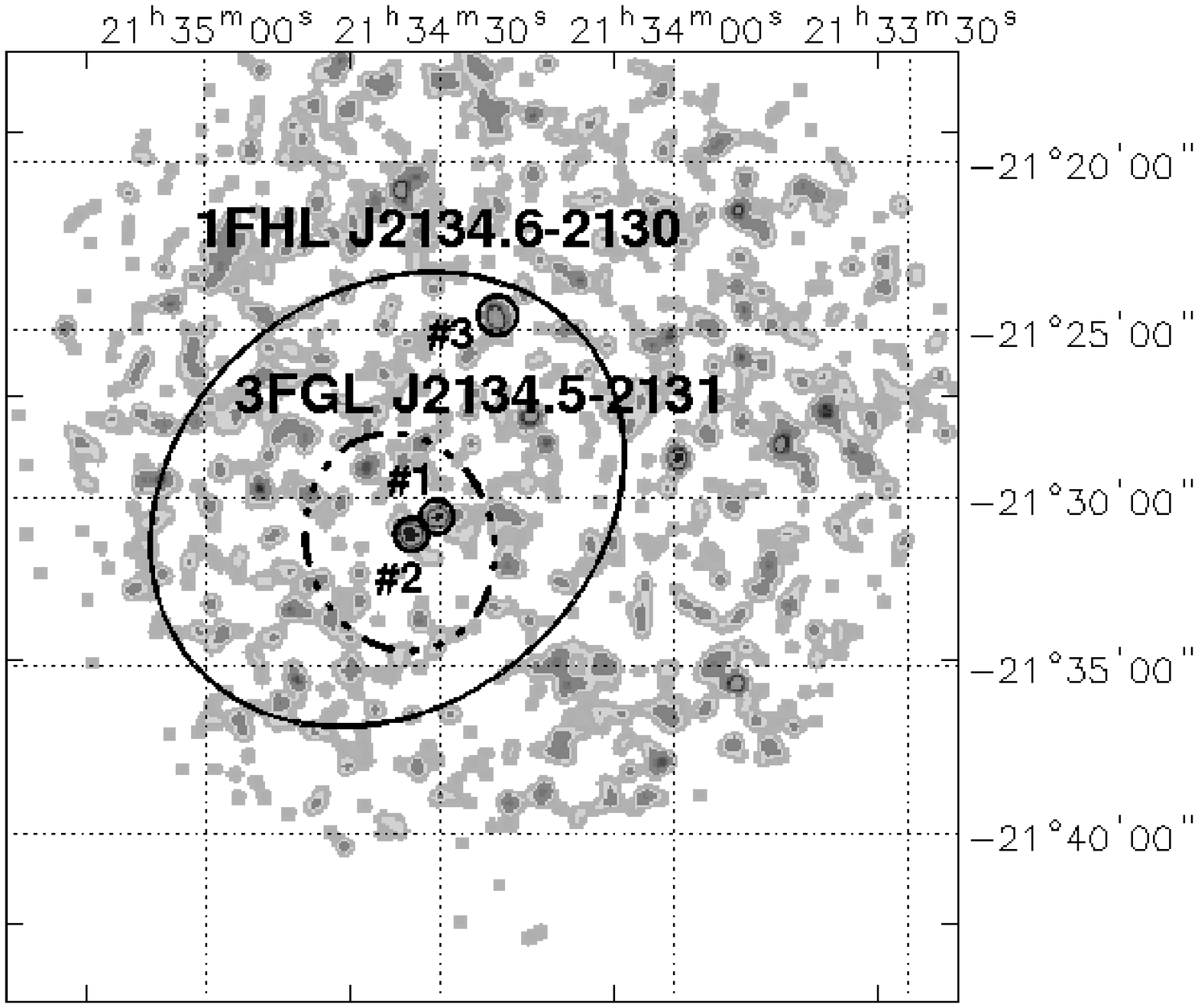}
\includegraphics[width=0.45\linewidth,angle=0]{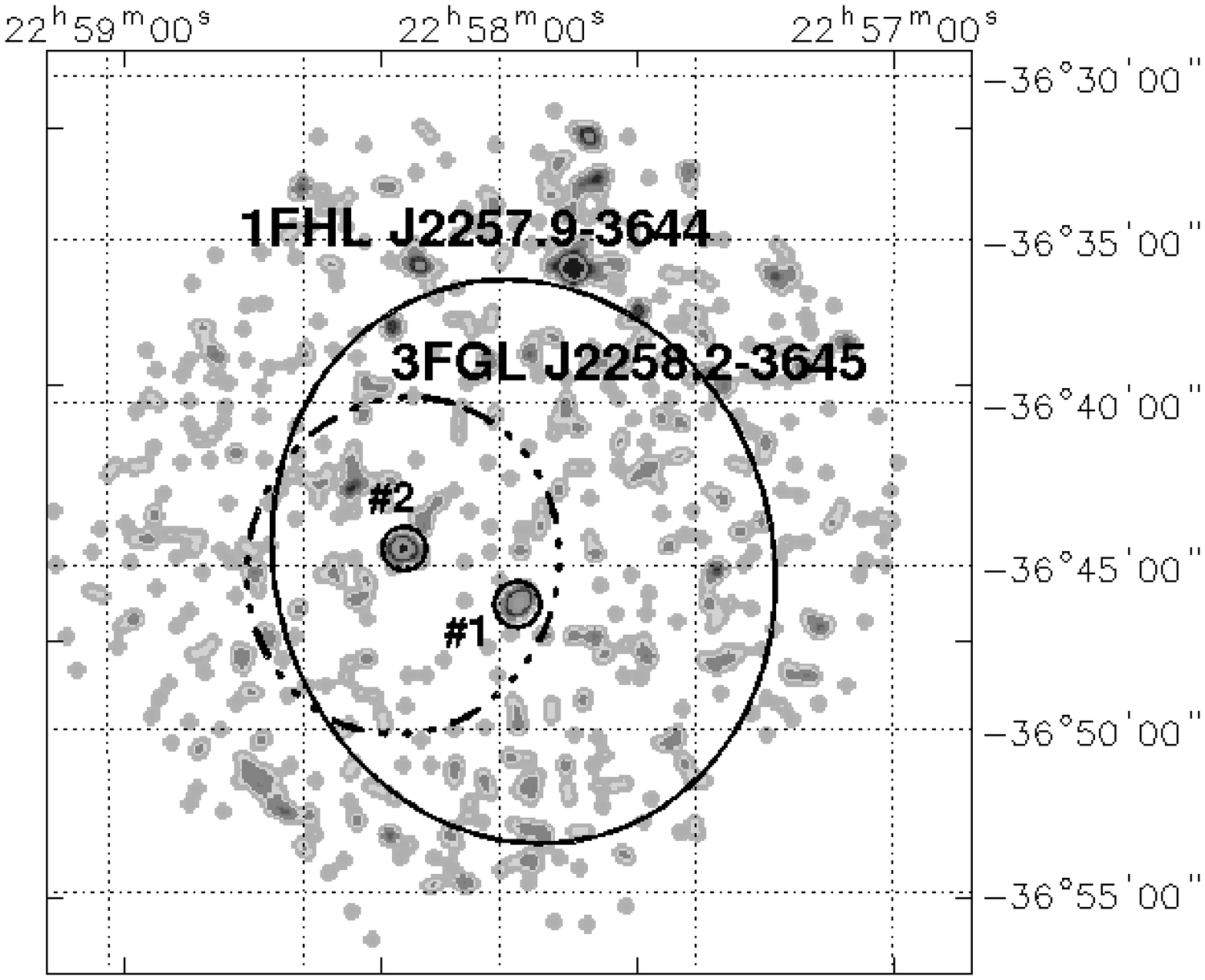}
\caption{XRT 0.3--10 keV images of high-energy \emph{Fermi} sources for which more than one X-ray 
detection 
is found. Black circles depict the location of each XRT detection. Black ellipse and black-dashed-dotted 
ellipse depict the positional uncertainty of the 1FHL and 3FGL sources, respectively.}
\label{f2}
\end{figure*}

\subsection{\textbf{Multiple counterpart sources}}

In the following cases, more than one single X-ray detection is found within the \emph{Fermi} error 
ellipse. Although this makes things more difficult to handle, it is still possible to highlight 
a more likely association on the basis of a multi-waveband approach.

\subsubsection{1FHL J0746.3--0225}

This is a case where there are two X-ray detections inside the \emph{Fermi} error ellipse (see upper left
panel of Figure~\ref{f2}), but only one (source \#1), which is the brightest, is a likely counterpart. It 
is in fact the only one with radio emission because it is listed in the NVSS survey as NVSS J074627--022549, 
with a 20 cm flux density of $\sim$12.8 mJy. It has a WISE counterpart with colours $W2-W3=2.1$ 
and $W1-W2=0.68$, as is
typical of gamma-ray blazars. This X-ray source is also UV-detected (GALEXASC J074627.06--022548.9) with 
far- and near-UV magnitude in the range 20.4--20.8. The other object is dimmer, has no interesting 
association in other catalogues, and is therefore a less likely association. 
In the latest \emph{Fermi} catalogue, 1FHL J0746.3--0225 is associated with 3FGL J0746.4--0225;
as is clear in Figure~\ref{f2}, the 
3FGL error ellipse confirms the association of the high-energy detection with source \#1 and excludes 
source \#2 as a possible counterpart.

\subsubsection{1FHL J1410.4+7408}

Also in this case we find a couple of detections that are compatible with the 1FHL positional uncertainty 
(see upper right panel of Figure~\ref{f2}). Both have WISE colours ($W2-W3=2.79$ and $W1-W2=0.88$ and 
$W2-W3=1.81$ and $W1-W2=0.79$ for \#1 and \#2, respectively), so just compatible with the blazar strip, 
but no detection in radio. We also note that within the 1FHL positional uncertainty, there is also a 
\emph{XMM-Newton} Slew source XMMSL1 J141002.6+740744 (5 arcsec error radius), which is quite bright in 
X-rays (0.2--12 keV flux of $2.3\times10^{-12}$ erg cm$^{-2}$ s$^{-1}$); despite this, we could not find 
a radio, infrared, or optical counterpart, even when doubling its positional uncertainty. Only going to 
an 18 arcsec distance are we able to find an association with a WISE object. Its WISE magnitudes are $W1= 
17.189$, $W2=16.771$, $W3=12.792$ and $W4=9.606$, and its colours turn out to be $W2-W3=3.98$ and 
$W1-W2=0.42$, which locate the source outside the blazar strip. Also this counterpart is not reported at 
radio frequencies. That this \emph{XMM-Newton} Slew source is detected in neither the combined XRT image 
nor the individual observations, indicates further that it is extremely variable in X-rays. Despite its 
potential interest, this \emph{XMM-Newton} Slew detection is, however, outside the error ellipse of the 
3FGL counterpart of this high-energy \emph{Fermi} source. The other two X-ray sources are instead both 
well inside the 3FGL positional uncertainty, so are more likely candidates. Clearly, this is another 
complicated case, and only optical follow-up observations can help to identify which of these two 
(possibly three) X-ray detections is the true counterpart to this high-energy emitter.

\subsubsection{1FHL J2134.6--2130}

In this case three XRT sources are found within the 1FHL error ellipse (see lower left panel of 
Figure~\ref{f2}). Source \#1 has a radio counterpart (NVSS J213430--213032) with a 20 cm flux 
density of $\sim$22 mJy and a flat radio index of 0.23 (Petrov et al. 2013). 
It is also reported in the list of possible compact radio counterparts to 2FGL sources 
by Schinzel et al. (2015), where it is described as a flat spectrum source with 6 and 3.3 cm 
flux densities of $\sim$92.8 and $\sim$103.9 mJy, respectively.
The WISE colours ($W2-W3=2.27$ and $W1-W2=0.78$) 
are just compatible with the blazar strip and the source is discussed by various authors as a possible 
counterpart to 2FGL J2134.6--2130; for example, Massaro et al. (2013b) and Hassan et al.(2013) suggest
that it could be a blazar of the BL Lac type. Source \#2 is listed as a galaxy in NED, but it has no 
detection in radio, and it does not display WISE colours typical of blazars, so we consider it an 
unlikely association. As for source \#3, the lack of any radio counterpart, as well as of WISE colours 
($W2-W3=3.213$ and $W1-W2=0.481$) that are not compatible with the blazar strip, suggests that this 
object is also
an unlikely association. The third \emph{Fermi} catalogue lists this object as 3FGL J2134.5--2131. Its 
restricted error ellipse excludes source \#3 as a likely counterpart, but maintains \#1 and \#2 as 
possible associations. On the basis of our analysis, we consider only the first to be the most promising 
counterpart to this \emph{Fermi} high-energy detection.

\subsubsection{1FHL J2257.9--3644}

Only two soft X-ray counterparts are found within the \emph{Fermi} error ellipse (see lower right panel of 
Figure~\ref{f2}). This 1FHL source is listed in the 3FGL catalogue as 3FGL J2258.2--3645. 
Even though its 
positional uncertainty is smaller than in the 1FHL survey, it does not help to exclude one of the two 
X-ray detections. Source \#1 is dim with no counterparts in radio catalogues and WISE colours 
$W2-W3=3.06$ and $W1-W2=1.46$, which locate the source at the upper border of the blazar strip. The 
other detection (source \#2 in the figure) is brighter and has a radio detection in the NVSS (NVSS 
J225815--364433) and in the SUMSS (SUMSS J225816--364446) with 20 and 36 cm flux densities of 
$\sim$10.6 and 
$\sim$14.1 mJy, respectively. The radio spectrum is quite flat displaying an index of 0.23 (Petrov et al. 
2013). The source is also reported in the WISE catalogue with colours $W2-W3=2.24$ and $W1-W2=0.57$, which 
are compatible with the blazar strip. It is also listed in the GALEX All-Sky survey (GALEXASC 
J225815.00--364434.6) with far- and near-UV magnitudes around 20--21. Source \#2 is reported as a galaxy 
in NED and has recently been classified as a BL Lac by Landoni et al. (2015). Optical spectroscopy of 
source \#1 could unambiguously exclude it as the counterpart of this \emph{Fermi} high-energy emitter.

\subsection{\textbf{Objects with X-ray detection outside the Fermi positional uncertainty}}

In two cases (1FHL J0639.6--1244 and 1FHL J1856.9+0252), we find a bright X-ray source outside the 
\emph{Fermi} error ellipse. Despite this, in each case the association may be considered at least 
interesting for a number of reasons: a) \emph{Fermi} error ellipses are quoted at the 95$\%$ confidence 
level, which suggests that in 5\% of the cases, the true counterpart could be located outside; b) the 
associations are with interesting objects, which may have the characteristics of gamma-ray 
emitting objects; c) there is nothing else inside the \emph{Fermi} positional uncertainty that could be of 
potential interest.

\begin{figure}[h]
\centering
\includegraphics[width=1.\linewidth,angle=0]{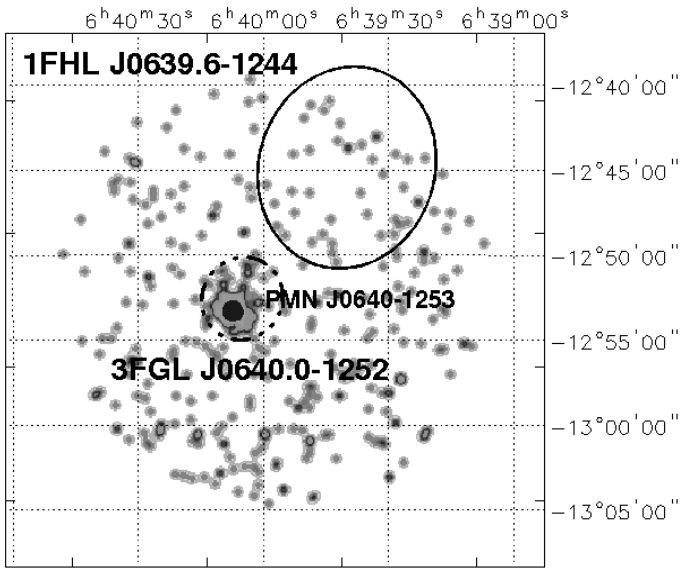}
\caption{XRT 0.3--10 keV image of the region surrounding 1FHL J0639.6--1244 (black ellipse). 
The only X-ray object detected by XRT, namely PMN J0640--1253, lies $\sim$5 arcmin away from the border of
\emph{Fermi} positional uncertainty, but is found to be within the error ellipse of 
3FGL J0640.0--1252 (black-dashed-dotted ellipse).} 
\label{f3}
\end{figure}

\subsubsection{1FHL J0639.6--1244}

The only X-ray detection lies around 11 arcmin away from the \emph{Fermi} position (see 
Figure~\ref{f3}). This object is associated with 2MASX J06400717--1253150 and with the flat spectrum radio 
object PMN J0640--1253 ($\alpha \sim -0.4$, Chhetri et al. 2013). This source is fairly bright in radio 
with a 20 
cm flux density of $\sim$225 mJy and is detected by WISE with colours ($W2-W3=1.74$ and $W1-W2=0.457$), 
which are once again fully compatible with the blazar strip. 
It is also quite bright in X-rays (see Table~\ref{tab4}), thus strengthening its association with 
1FHL J0639.6--1244.

It was proposed as a BL Lac candidate for TeV emission by 
Massaro et al. (2013b). The nearest 3FGL source is 3FGL J0640.0--1253, but the two \emph{Fermi} detections 
are not associated in the 3FGL catalogue, given that their positional uncertainties do not overlap (see 
Figure~\ref{f3}). As is evident from the figure, the 3FGL source is associated with PMN J0640--1253.
Seeing as there is no other X-ray emitter in this zone and that the 3FGL source could be expected to 
radiate even above 10 GeV, it is possible that the 1FHL error circle is slightly underestimated and then 
the two \emph{Fermi} objects could well be the same source. We therefore consider the association 
possible.

\subsubsection{1FHL J1856.9+0252}

\begin{figure}[h]
\centering
\includegraphics[width=1.\linewidth,angle=0]{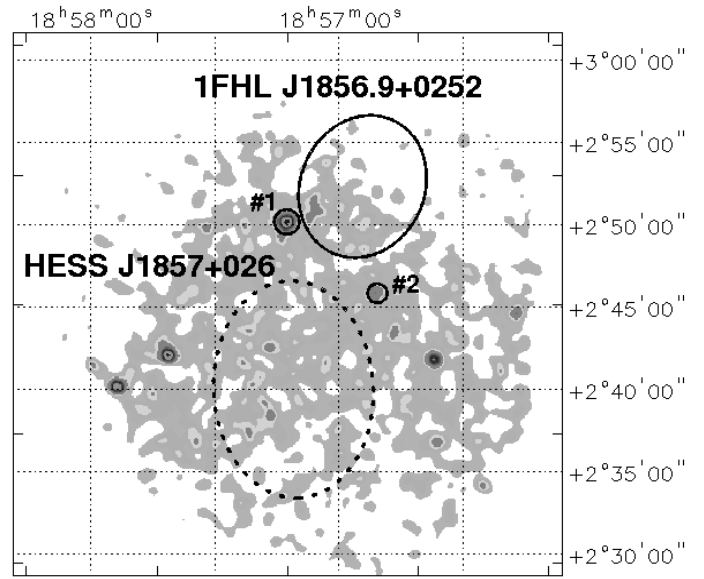}
\caption{XRT 0.3--10 keV image of the region surrounding 1FHL J1856.9+0252 
(black ellipse). Source \#1 and \#2 are the sources detected by XRT in the region surrounding 
the \emph{Fermi} object. Source \#2 coincides with the pulsar PSR J1856+0245, 
which has been suggested to power the HESS object (black-dotted ellipse).}
\label{f4}
\end{figure}

This source is not reported in the third \emph{Fermi} catalogue. As displayed in Figure~\ref{f4}, it is 
not far from the TeV emitter HESS J1857+026, which is generally classified as a PWN candidate (Acero et 
al. 2013); the morphology and overall spectral shape of the main emission zone, combined with the 
proximity to the pulsar PSR J1856+0245 (Hessels et al. 2008), support this classification.

Unfortunately, the two error ellipses (from \emph{Fermi} and from TeV observations) do not overlap, which 
makes the possible association between the two wavebands uncertain. Nearby the \emph{Fermi} error ellipse 
XRT detects source \#1 whose WISE colours ($W2-W3=1.91$ and $W1-W2=1.71$) locate it well outside the 
blazar strip; furthermore, this object has no counterpart in radio surveys. A second source (\#2 in 
Figure~\ref{f4}) coincides with the pulsar PSR J1856+0245, which is thought to be responsible for the HESS 
emission. It is difficult at this stage to conclude which of the two sources is the true association 
with the 1FHL object, nor is it clear whether the HESS and \emph{Fermi} detection are the same object, 
leaving this case unresolved. For the following, we consider source \#1 as the possible association.

\begin{figure*}[t]
\centering
\includegraphics[width=12cm,height=9cm,angle=0]{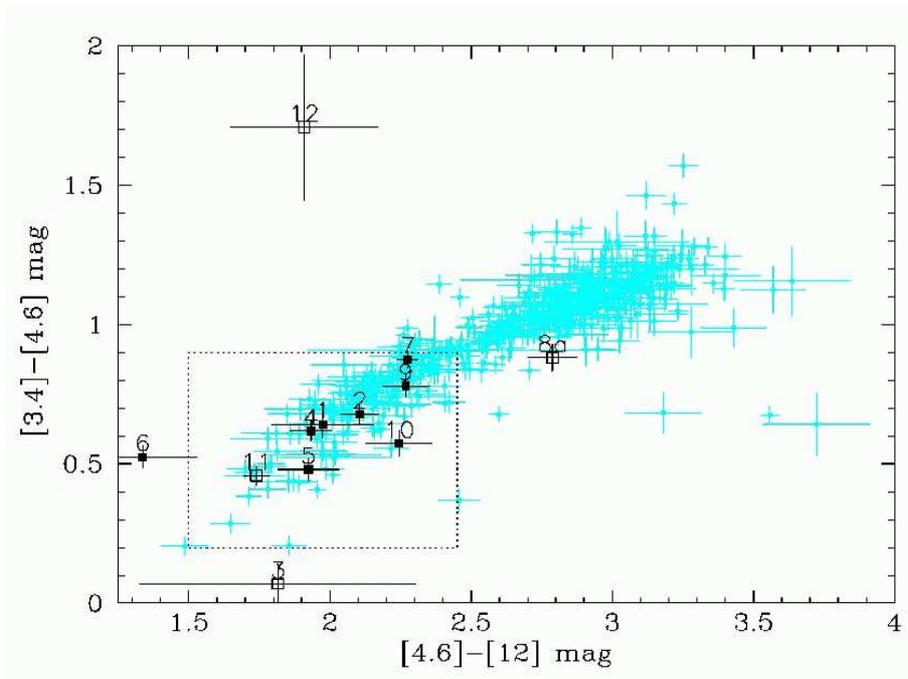}
\caption{[4.6]--[12]/[3.4]--[4.6] MIR colour-colour plot reporting the positions of
gamma-ray emitting blazars (in cyan) associated with WISE
sources forming the blazar strip (see Massaro et al. (2013a) for more details), together with the
BL Lac objects identified in this paper. Filled and unfilled squares depict the proposed associations 
flagged with \emph{L} or \emph{P} in the last column of Table~\ref{tab3}, respectively.}
\label{f5}
\end{figure*}

\section{Discussion and conclusion}

The first result of this work is that we have uncovered a number of likely X-ray counterparts to still
unassociated 1FHL gamma-ray sources. The majority of these associations are likely to be 
extragalactic in nature, most probably blazars of some type. Only in the case of 1FHL J1115.0--0701
is the association considered to be unlikely. 
In four other cases, the proposed associations are to be taken as possible.
More specifically, for 1FHL J1115.0--0701 and 1FHL J1410.4+7408, the X-ray detections are compatible with 
the 1FHL positional uncertainty, but the overall multi-wavelength properties prevent us from drawing any 
firm conclusions about their nature and only future optical follow-up observations can help to provide a 
reliable classification of these X-ray sources.
In the other two cases (1FHL J0639.6--1244 and 1FHL J1856.9+0252), the X-ray detection lies outside the 
1FHL error ellipse.
For 1FHL J0639.6--1244, there is only one bright X-ray source detected in the region surrounding the 1FHL 
source, which coincides with PMN J0640--1253, a flat spectrum radio source that has been recently 
associated with the third \emph{Fermi} catalogue source 3FGL J0640.0--1252. Although the 1FHL and 3FGL 
error ellipses do not overlap, we suggest that a likely association between PMN J0640--1253 and the 1FHL 
source cannot be ruled out.  
Also in the case of 1FHL J1856.9+0252 the overall multi-waveband properties of the only XRT detection 
do not allow us to associate firmly the X-ray source with the 1FHL object.

In the remaining cases, at least one of the proposed associations has a radio detection and WISE colours 
compatible with the blazar strip, in other words, it qualifies as beamed AGN. Indeed two objects (1FHL 
J1223.3+7953 and 1FHL J2257.9--3644) have been optically classified as BL Lac objects. As already noted by 
Stephen et al. (2010) and confirmed by Masetti et al. (2013), the association of \emph{Fermi} sources with 
soft X-ray counterparts favours discovery of this type of blazar. The preference for finding BL Lac 
when using soft X-ray data is very likely related to the SED of these objects compared to flat spectrum radio 
quasars. In fact, the X-ray selection favours the discovery of high synchrotron peaked or HSP blazars, 
which are also good candidates for TeV emission, because the Compton peak in these AGN is expected in this 
energy range owing to the presence of high-energy electrons.
 
As discussed by Masetti et al. (2013) and in the references therein, these objects populate a well-defined 
region 
of the WISE colour-colour diagram, i.e a square located in the lower part of the blazar strip. In fact, if 
we plot in the WISE colour-colour diagram (see Figure~\ref{f5}), the most likely counterparts to 1FHL 
sources listed in Table~\ref{tab3} (i.e. considering only those sources flagged with \emph{L} or 
\emph{P} in the last column of the Table), we notice that most fall along the blazar strip or close 
to it 
but, most important, nine of them lie within the limits of the locus populated by TeV-emitting BL Lacs,
while one is at its border. Of the two possible counterparts of 1FHL J1410.4+7408, one lies 
within the TeV square (8b), while the other is on the blazar strip (8a), thus suggesting that both are 
likely associations. Finally, Figure~\ref{f5} confirms that the association proposed here for 1FHL 
J1856.9+0252 (\#12 in the figure) remains highly uncertain. We therefore conclude once again that 
the association among 
unassociated high-energy \emph{Fermi} sources and soft X-ray objects appears to select 
blazars, for a large percentage of the HSP BL Lac type, i.e. those that are good candidates for TeV 
emission. Clearly, only
the optical spectroscopy of all soft X-ray counterparts discussed in this work can confirm this suggestion 
and 
provide further insight into gamma-ray selected blazars in general and BL Lac in particular.

\begin{acknowledgements} 

This research made use of  the  NASA/IPAC Extragalactic Database (NED) operated by the
Jet Propulsion Laboratory (California Institute of Technology) and of the HEASARC 
archive provided by NASA's Goddard Space Flight Center. The authors also acknowledge the use of public 
data from the \emph{Swift} data archive.
R. L. acknowledges financial support under contract INTEGRAL ASI I/033/10/0.
N. M. thanks Francesco Massaro for help with the preparation of Figure 5.
\end{acknowledgements}

\clearpage

\begin{table*}
\begin{minipage}{155mm}
\centering
\caption{Log of the \emph{Swift}/XRT observations used in this paper (until March 15, 2015).}
\label{tab1}
\begin{tabular}{lccc}
\hline
\hline
\emph{Fermi} source  &     ID        &   Obs date     & Exposure       \\
                     &               &                &      (s)        \\
\hline
\hline
1FHL J0312.8+2013    &  00047145001  &  Jul 01, 2012  &  2186  \\
                     &  00047145002  &  Jul 05, 2012  &  1645  \\
                     &  00047145003  &  Jul 07, 2012  &  281   \\
total obs            &     --        &      --        &  4112  \\
\hline
1FHL J0625.9+0002    &  00041322001  &  Aug 23, 2010  &  2516  \\
                     &  00041322002  &  Sep 08, 2010  &  1868  \\
total obs            &      --       &      --        &  4384  \\
\hline
1FHL J0639.6--1244   &  00083670001  &  Feb 27, 2014  &  711   \\
\hline
1FHL J0644.2+6036    &  00047166002  &  Feb 09, 2012  &  1494  \\
                     &  00047166003  &  Feb 13, 2012  &  1762  \\
total obs            &     --        &      --        &  3256  \\
\hline
1FHL J0746.3--0225   &  00047174002  &  Feb 25, 2012  &  2940  \\
                     &  00047174003  &  Nov 05, 2012  &  1408  \\
total obs            &      --       &        --      &  4348  \\
\hline
1FHL J0928.1--5252   &  00084691001  &  Feb 05, 2015  &  116   \\
                     &  00084691002  &  Feb 09, 2015  &  1597  \\
                     &  00084691003  &  Feb 15, 2015  &  133   \\
                     &  00084691004  &  Apr 03, 2015  &  83    \\
total obs            &       --      &       --       &  1929  \\
\hline
1FHL J1115.0--0701   &  00047198001  &  Jan 24, 2012  &  718   \\
                     &  00047198002  &  Mar 20, 2012  &  699   \\
                     &  00047198004  &  Mar 29, 2012  &  672   \\
                     &  00047198006  &  Jun 22, 2012  &  1296  \\
total obs            &     --        &      --        &  3385  \\
\hline
1FHL J1129.2-7759    &  00084723001  &  Feb 03, 2015  &  276   \\
                     &  00084723002  &  Feb 04, 2015  &  231   \\
                     &  00084723003  &  Feb 05, 2015  &  448   \\
                     &  00084723004  &  Feb 07, 2015  &  293   \\
                     &  00084723005  &  Feb 08, 2015  &  905   \\
                     &  00084723006  &  Feb 15, 2015  &  662   \\
                     &  00084723007  &  Mar 06, 2015  &  602   \\
total obs            &       --      &       --       &  3417  \\
\hline
1FHL J1223.3+7953    &  00047201001  &  Feb 10, 2012  &  4179   \\
\hline
1FHL J1240.4--7150   &  00041386001  &  Feb 12, 2011  &  3161   \\
                     &  00041386002  &  Feb 14, 2011  &  1605   \\
total obs            &      --       &      --        &  4766   \\
\hline  
1FHL J1315.7--0730   &  00041395001  &  Aug 18, 2010  &  4958    \\
                     &  00041395002  &  Aug 21, 2010  &  1678    \\
total obs            &      --       &       --       &  6636    \\
\hline  
1FHL J1407.1--6133   &  00042321001  &  Mar 10, 2012  &  617    \\
\hline
1FHL J1410.4+7408    &  00041402001  &  Mar 08, 2011  &  1414   \\
                     &  00041402003  &  Mar 11, 2011  &  2007   \\
                     &  00047219004  &  Mar 29, 2012  &  1026   \\
                     &  00047219006  &  Apr 05, 2012  &  1171   \\
                     &  00084043009  &  Jun 12, 2014  &  1258   \\ 
total obs            &      --       &       --       &  6876    \\
\hline
1FHL J1507.0--6223   &  00047226002  &  Dec 18, 2012  &  2009    \\
                     &  00047226003  &  Dec 19, 2012  &  1826    \\
total obs            &      --       &       --       &  3835    \\
\hline
1FHL J1619.8+7540    &  00084770001  &  Jan 23, 2015  &  756    \\
\hline
1FHL J1634.7--4705   &  00042964002  &  Jan 16, 2012  &  341    \\
                     &  00410087000  &  Jan 29, 2012  &  516    \\ 
                     &  00042957001  &  Apr 28, 2012  &  543    \\ 
                     &  00080704001  &  Mar 12, 2014  &  788   \\
                     &  00080705001  &  Mar 12, 2014  &  2059   \\ 
total obs            &      --       &       --       &  4247    \\
\hline
1FHL J1758.3--2340   &  00049691001  &  Jun 28, 2014  &  203    \\
                     &  00043785001  &  Oct 05, 2012  &  513    \\
total obs            &      --       &       --       &  716    \\
\hline
1FHL J1839.1--0557   &  00044397001  &  Mar 15, 2013  &  527    \\ 
\hline
1FHL J1839.4--0708   &  00044371001  &  Mar 15, 2013  &  528    \\
                     &               &                &     \\
                     &               &                &     \\
\end{tabular}
\end{minipage}
\end{table*}
 \begin{table*}
 \begin{minipage}{155mm}
 \centering
 \begin{tabular}{lccc}
 \hline
 \hline
Source               &     ID        &   Obs date     & Exposure        \\
                     &               &                &      (s)        \\ 
\hline
\hline
1FHL J1856.9+0252    &  00036183001  &  Nov 10, 2006  &  720   \\  
                     &  00036184001  &  Mar 01, 2007  &  3097  \\
                     &  00036184002  &  Mar 07, 2007  &  4107  \\ 
                     &  00036183002  &  Mar 13, 2007  &  4102  \\ 
                     &  00037742001  &  Nov 16, 2008  &  9148  \\ 
                     &  00044642002  &  Mar 02, 2013  &  456   \\
                     &  00032774001  &  Mar 27, 2013  &  782   \\  
                     &  00032782001  &  Mar 28, 2013  &  782   \\
total obs            &      --       &       --       &  23194  \\
\hline
1FHL J2004.4+3339    &  00041467001  &  Sep 05, 2010  &  448   \\  
                     &  00041467002  &  Jan 09, 2011  &  1605  \\
                     &  00041467004  &  Jan 16, 2011  &  2040  \\ 
total obs            &      --       &       --       &  4093  \\ 
\hline
1FHL J2134.6--2130   &  00041496001  &  Sep 26, 2010  &  439   \\
                     &  00041496002  &  Dec 19, 2010  &  1838  \\  
                     &  00041496003  &  Dec 23, 2010  &  4274  \\
                     &  00084044001  &  Jul 27, 2014  &  1304  \\ 
                     &  00084044002  &  Aug 05, 2014  &  1036  \\
                     &  00084044003  &  Sep 12, 2014  &  156   \\  
                     &  00084044004  &  Sep 14, 2014  &  453   \\
                     &  00084044005  &  Oct 13, 2014  &  739   \\ 
                     &  00084044006  &  Oct 15, 2014  &  273   \\ 
                     &  00084044008  &  Dec 18, 2014  &  1203  \\
total obs            &      --       &       --       &  11715  \\
\hline                     
1FHL J2257.9--3644   &  00041507001  &  Aug 02, 2010  &  1734   \\
                     &  00041507002  &  Aug 12, 2010  &  1800   \\
total obs            &      --       &       --       &  3534   \\
\hline
\hline
\end{tabular}
\end{minipage}
\end{table*}

\clearpage

\begin{table*}
\begin{center}
\caption{High-energy \emph{Fermi} sources for which we do not find any X-ray counterpart.}
\begin{tabular}{lcc}
\hline
\hline
\emph{Fermi} source             &  Count rate$^{\dagger}$ &     $N_{\rm{H(Gal)}}$  \\
                                & ($10^{-3}$ counts s$^{-1}$) &  ($10^{22}$ cm$^{-2}$)  \\
\hline
\hline
1FHL J0312.8+2013               &           $<2.71$              &    0.0988   \\
1FHL J0625.9+0002               &           $<1.80$              &    0.413    \\
1FHL J0928.1--5252              &           $<4.00$              &    0.938    \\
1FHL J1407.1--6133              &           $<14.7$              &    2.00     \\
1FHL J1619.8+7540               &           $<0.67$              &    0.0382   \\
1FHL J1634.7--4705              &           $<0.65$              &    1.68     \\
1FHL J1758.3--2340              &           $<0.44$              &    1.12     \\
1FHL J1839.1--0557$^{\ddagger}$ &           $<0.59$              &    1.74     \\
1FHL J1839.4--0708              &           $<0.39$              &    1.44     \\
1FHL J2004.4+3339               &           $<0.98$              &    1.15      \\ 
\hline
\hline
\label{tab2}
\end{tabular}
\end{center}
\begin{list}{}{}
\item $^{\dagger}$ Count rates are extracted in the 0.3--10 keV energy band;
\item $^{\ddagger}$ Source in the field of HESS J1841--055.
\end{list}
\end{table*}

\begin{landscape}
\begin{table}
\centering
\caption{List of the XRT detections and their WISE properties found for each high-energy \emph{Fermi} 
source.}
\begin{tabular}{lcccccccccc}
\hline
\hline
\multicolumn{1}{c}{\emph{Fermi} source}   & \multicolumn{4}{c}{XRT source$^{\dagger}$} & 
\multicolumn{5}{c}{WISE counterpart} & \multicolumn{1}{c}{Association$^{\ddagger}$} \\

        & Source  & R.A.  & Dec. & error &  \multicolumn{1}{c}{Name} & 
\multicolumn{4}{c}{Magnitudes}  &  \\
        &   & (J2000) & (J2000) & (arcsec) & & 
 W1[3.4 $\mu$m] & W2[4.6 $\mu$m] & W3[12 $\mu$m] & W4[22 $\mu$m] &   \\
\hline
\hline
1FHL J0644.2+6036   & single(1) & 06 44 36.48 &   +60 38 49.40 & 4.2  & WISE J064435.72+603851.2   & 14.272 
& 13.631 & 11.657 & 9.246   & \textbf{L}\\
\hline
1FHL J0746.3--0225  & \#1(2)    & 07 46 27.14 &  --02 25 50.70 & 3.7  & WISE J074627.03--022549.3  & 13.090 
& 12.413 & 10.308 & 8.314   & \textbf{L}\\
                    & \#2    & 07 45 54.93 &  --02 24 31.40 & 4.4  & WISE J074554.80--022430.7  & 15.559 
& 14.501 & 11.974 & 8.968   & U\\
\hline
1FHL J1115.0--0701  & single(3) & 11 15 15.30 &  --07 01 26.00 & 6.0  & WISE J111515.34--070125.7  & 14.428 
& 14.359 & 12.534 & 8.561   & \textbf{P}\\   
\hline
1FHL J1129.2-7759   & single(5) & 11 30 32.25 &  --78 01 05.20 & 3.6  & WISE J113031.99--780105.5  & 13.323 
& 12.705 & 10.772 & 8.828   & \textbf{L} \\
\hline   
1FHL J1223.3+7953   & single(4) & 12 23 59.71 &   +79 53 24.40 & 5.1  & WISE J122358.17+795327.8   & 13.871 
& 13.391 & 11.467 & 9.236   & \textbf{L}\\
\hline
1FHL J1240.4--7150  & single(6) & 12 40 21.34 &  --71 48 58.51 & 3.6  & WISE J124021.21--714857.7  & 13.396 
& 12.871 & 11.534 &  8.648  & \textbf{L} \\ 
\hline
1FHL J1315.7--0730  & single(7) & 13 15 53.06 &  --07 33 01.80 & 3.6  & WISE J131552.98--073301.9  & 12.371 
& 11.496 & 9.221  & 7.185   & \textbf{L}\\
\hline
1FHL J1410.4+7408   & \#1(8a)  &   14 10 52.10  & +74 04 14.40   & 6.0  & WISE J141052.02+740415.1  & 14.872   
& 13.991   &11.203  & 9.029 & \textbf{P}\\
                    & \#2(8b)  &   14 10 45.40  & +74 05 10.40   & 6.0  & WISE J141046.01+740511.2  & 14.850   
& 14.051   &12.233  & 9.153 & \textbf{P}\\
\hline
1FHL J1507.0--6223  & single & 15 07 59.00 &  --62 25 21.80 & 6.0  & WISE J150758.80--622526.9 & 12.187 & 12.198 & 11.027 & 8.254     & U \\
\hline
1FHL J2134.6--2130  & \#1(9)    & 21 34 30.40 &  --21 30 33.00 & 5.0 &  WISE J213430.18--213032.6  & 13.548 
& 12.768 & 10.499 & 8.717   & \textbf{L}\\
                    & \#2    & 21 34 33.31 &  --21 31 05.30 & 5.6 &  WISE J213433.41--213103.2  & 15.238 & 14.728 & 11.582 & 8.725   & U\\
                    & \#3    & 21 34 22.60 &  --21 24 33.50 & 6.0 &  WISE J213423.04--212435.6  & 16.179 & 15.698 & 12.485 & 9.027   & U\\
\hline
1FHL J2257.9--3644  & \#1    & 22 57 57.38 &  --36 46 11.90 & 4.9 &  WISE J225757.06--364608.5  & 
15.342 & 13.880 & 10.824 & 8.520   & U \\
                    & \#2(10)    & 22 58 14.74 &  --36 44 28.80 & 5.1 &  WISE J225815.00--364434.2  & 
13.860 & 13.287 & 11.042 & 8.916   & \textbf{L}\\
\hline
\hline
\multicolumn{11}{c}{Objects with X-ray detection outside the \emph{Fermi} positional uncertainty}\\
\hline
\hline
1FHL J0639.6--1244  & single(11) & 06 40 07.31 &  --12 53 18.60 & 3.8  &   WISE J064007.19--125315.0  & 
11.891 & 11.434 & 9.694 & 7.745 & \textbf{P} \\
\hline
1FHL J1856.9+0252   & single(12) & 18 57 12.70 &  +02 50 11.30 & 3.3 &   WISE J185712.64+025007.8   & 14.253 
& 12.545 & 10.637 & 7.299   & \textbf{P}\\ 
\hline
\hline
\label{tab3}
\end{tabular}
\begin{list}{}{}
\item $^{\dagger}$ The number in parenthesis next to the XRT sources flagged either with \emph{L} or 
\emph{P} in the last column of the Table is a reference used in the WISE colour-colour plot (see
Figure~\ref{f5} to identify every specific source);
\item $^{\ddagger}$ \emph{L} = likely, \emph{P}= possible, \emph{U} = unlikely.
\end{list}
\end{table} 
\end{landscape}

\begin{table*}
\centering
\caption{XRT data analysis results.}
\begin{tabular}{lcccccccccc}
\hline
\hline
\multicolumn{1}{c}{\emph{Fermi} source}   & \multicolumn{5}{c}{XRT source} \\

   & Source  & Count rate  & $N_{\rm{H(Gal)}}$ & $\Gamma$ & $F(2-10~\rm{keV})$   \\
   &         & ($10^{-3}$ cts s$^{-1}$) & ($10^{22}$ cm$^{-2}$) & &($10^{-12}$ erg cm$^{-2}$ s$^{-1}$) \\
\hline
\hline
1FHL J0639.6--1244  & single &  $430.10\pm24.67$ &  0.299   &  $2.07\pm0.17$           &   13.1   \\ 
1FHL J0644.2+6036   & single &  $16.45\pm0.49$   &  0.065   &  $2.38^{+0.51}_{-0.56}$  &   0.26   \\
\hline
1FHL J0746.3--0225  & \#1    &  $28.48\pm2.63$   &  0.0769  &  $2.64\pm0.37$           &   0.37   \\
                    & \#2    &  $5.03\pm1.30$    &  0.0796  &       --                 &    --     \\
\hline
1FHL J1115.0--0701  & single &  $3.08\pm1.10$    &  0.0348  &       --                 &    --     \\ 
\hline
1FHL J1129.2-7759   & single &  $331.10\pm10.05$ &  0.073   &  $1.96\pm0.08$           &    7.62   \\  
\hline   
1FHL J1223.3+7953   & single &  $5.71\pm1.40$    &  0.052   &       --                 &    --     \\
\hline
1FHL J1240.4--7150  & single &  $279.70\pm7.71$  &  0.140   &  $1.96\pm0.07$           &    7.76   \\
\hline
1FHL J1315.7--0730  & single &  $63.56\pm3.14$   &  0.0308  &  $2.68\pm0.13$           &    0.56    \\
\hline
1FHL J1410.4+7408   & \#1    &  $1.57\pm0.57$    &  0.0223  &       --                 &     --     \\
                    & \#2    &  $2.84\pm0.79$    &  0.0225  &       --                 &     --     \\  
\hline
1FHL J1507.0--6223  & single &  $3.11\pm1.10$    &  0.419   &       --                 &     --     \\ 
\hline
1FHL J1856.9+0252$^{\dagger}$   & single &  $2.80\pm0.38$    &  1.510   &      [1.8]               &     
0.50    \\ 
\hline
1FHL J2134.6--2130  & \#1    &  $1.83\pm0.49$    &  0.0333  &       --                 &      --     \\
                    & \#2    &  $2.19\pm0.53$    &  0.0334  &       --                 &      --     \\
                    & \#3    &  $1.18\pm0.42$    &  0.0325  &       --                 &      --     \\
\hline
1FHL J2257.9--3644  & \#1    &  $5.33\pm1.50$    &  0.0114  &       --                 &      --     \\   
                    & \#2    &  $6.10\pm1.50$    &  0.0115  &       --                 &      --     \\
\hline
\hline
\label{tab4}
\end{tabular}
\begin{list}{}{}
\item $^{\dagger}$ Despite of the poor statistical quality of the XRT data, the X-ray data indicate the 
presence of an intrinsic absorption $N_{\rm{H(intr)}} = \left(21.2^{+12.3}_{-8.9}\right)\times10^{22}$ 
cm$^{-2}$.
\end{list}
\end{table*}

\end{document}